\documentclass[12pt,preprint]{aastex}

\shortauthors{Watson \& Wyld}
\begin{document}
\title{Apparent Sizes and Spectral Line Profiles for Spherical and Disk Masers: Solutions to the Full Equations}

\author{W.D. Watson and H.W. Wyld}
\affil{Department of Physics, University of Illinois, 1110 West Green Street, Urbana, IL 61801} 
\email{w-watson@uiuc.edu and hwwyld@uiuc.edu}


\begin{abstract}
Calculations are performed for the spectral line profiles and images of astrophysical maser radiation that emerges from isolated spheres and thin disks viewed edge-on. In contrast to previous investigations where various approximations are made, the full equations are solved here for the frequency-dependent radiative transport that includes the thermal motion of the molecules. The spectral line profiles for spheres and disks are found to rebroaden to the full thermal Doppler breadth with increasing saturation in essentially the same way as is well known to occur for a linear maser. The variation with frequency in the apparent angular sizes of masing spheres and thin disks is found to be negligible at frequencies within the spectral line where the flux is significant. Calculations also are performed for spherical and disk  masers that are not isolated, but for which the seed radiation for the masers is incident from one side as would occur when a strong continuum source is on the far side of the masers. Again, the spectral line profiles are found to rebroaden to the full thermal breadths with increasing saturation and there are no significant variations in the apparent angular sizes with frequency.  However, the full rebroadening does occur at somewhat higher saturation and the variation of the apparent angular sizes as a function of the degree of saturation is quite different from that of the isolated masers. Spheres and disks have served as idealized geometries with which to examine possible deviations from the linear approximation for astrophysical masers.
\end{abstract}


\section{INTRODUCTION}

Interpretations of astrophysical masers which involve the radiative transfer of the maser radiation are based almost entirely on the ``linear maser'' (i.e., the one dimensional) approximation. In this approximation, the paths of the rays of radiation within the maser are parallel, straight lines. This approximation is motivated by the recognition that the longest maser rays tend to dominate as a result of the exponential gain in the intensity with path length in the unsaturated regime. From an observational viewpoint, that the rays are nearly parallel is indicated by evidence that the angles into which the emergent maser radiation is beamed are quite small---as small as 10$^{-5}$ ster in some cases (Nedoluha \& Watson 1991).

As an alternative to the extreme approximation that the masers are purely linear, the transfer of radiation in isolated spherical masers has been treated in a few investigations. Although it is generally recognized that astrophysical masers are unlikely to be ideal spheres, spheres nevertheless serve as a convenient geometry with which to assess possible deviations from the linear maser approximation. The initial studies (Litvak 1971; Goldreich \& Keeley 1972) demonstrated that the maser radiation becomes tightly beamed along the radii so that the rays are again nearly parallel. Consequently, with increasing saturation the apparent sizes of these masers as seen by a distant observer become much smaller than the geometrical sizes. In these studies, the maser radiation was treated as containing only a single frequency
and the thermal (and any other) velocities of the molecules were ignored.  Efforts to treat the full spectrum of the maser radiation and the thermal motions of the molecules in a spherical maser seems to have begun with Bettweiser \& Kegel (1974). The key aspect of the radiative transfer for masers that are not completely linear was recognized---the frequency dependence of the amplification of a ray varies with its direction of propagation. For spheres, the amplification depends upon the angle between the ray and the radius vector. Bettweiser (1976) concluded that the observable properties of spherical masers, such as the spectral line breadths and the apparent sizes as a function of frequency, may be sensitive to the details of the
treatment of the frequency-dependent radiative transport. However, with limited computer capabilities, definite conclusions could not be reached in that investigation. To simplify the computations somewhat, Emmering \&\ Watson (1994; hereafter EW) then developed a series expansion method that is based on the knowledge that the radiation is beamed into small angles about the radii.  Despite the complications in the detailed treatment of the anisotropic, frequency dependent radiative transport, on the issues of main interest EW found (a) that the spectral line narrows and rebroadens as a function of the degree of saturation in essentially the same way as for a linear maser, and (b) that the variation in the apparent angular size as a function of frequency is negligible at frequencies within the spectral line where the flux is
appreciable. Speculation to the contrary on these issues has nevertheless persisted--- even including assertions of a specific behavior that the line profiles from disks and spheres rebroaden only to 1/$\sqrt{2}$ and 1/$\sqrt {3}$ of the thermal Doppler breadth, respectively, instead of rebroadening to the full Doppler breadth. The calculation of EW does employ an involved
expansion proceedure and some reasoning is required to recognize that the results are accurate. The reasoning may not be completely clear to all readers. To remove these challenges to understanding that the key conclusions are valid, we have now computed the spectra and images by solving the full equations without approximations. In addition to spherical masers, masers in
the shape of thin disks and viewed edge-on also are treated. Disk masers are, perhaps, more likely than spherical masers to occur in the astronomical environment.  The results from the calculations with the full equations are described in this Paper.

The issues under investigation here are relevant for interpreting observational data. The observed breadths and shapes of the spectral lines of astrophysical masers can be used to infer information about the temperature, the degree of saturation, and the beaming angle of masers (e.g., Nedoluha \& Watson 1991; Watson, Sarma, \& Singleton 2002) if the theoretical behavior of
the profile is understood. It is striking that the spectral linebreadths associated with some of the brightest astrophysical masers (the H$_{2}$O maser flares) are much less than the Doppler breadths for the temperatures ($\ga$ 400K) that are expected for the gas in the relevant environments. The images of astrophysical masers often are not identical when observed at different frequencies within the spectral line (e.g., Gwinn 1994; Moscadelli et al. 2003). Establishing whether this difference is, at least in part, a natural consequence of the radiative transfer is necessary when considering its significance.

The relevant equations to be solved are given in Section 2. Results of computations for the spectral line breadth and for the apparent angular size of the masers are presented in Section 3. As in the previous investigations, the focus in the above discussion is on ``isolated'' masers---that is, on masers for which the seed radiation is due either to spontaneous emission or to isotropic continuum radiation that is incident from outside the masing region. In practice, some masers in the astronomical environment may be close enough to HII regions or other strong sources of continuum radiation that the radiation from one side dominates as the source of the seed radiation.  Calculations are performed in Section 4 for this case, as well, and compared with the spectral line breadths and apparent angular sizes obtained for isolated spherical and disk masers. 

\section{BASIC EQUATIONS}

For simplicity, we consider a masing transition between two energy levels
which have statistical weights of one. Let $n_{i}(\overrightarrow
{r},\overrightarrow{v})d\overrightarrow{r}d\overrightarrow{v}$ be the number
of atoms of level $i$ with velocities in the range $(\overrightarrow
{v},\overrightarrow{v}+d\overrightarrow{v})$ and located in the volume element
$d\overrightarrow{r}$ at position $\overrightarrow{r}$. The rate equations for
the populations of the upper $u$ and lower $l$ levels can then be written as%
\begin{equation}
0=\lambda_{u}(v)-\Gamma n_{u}(\overrightarrow{r},\overrightarrow{v}%
)-n_{-}(\overrightarrow{r},\overrightarrow{v})BJ_{k}(\overrightarrow
{r},\overrightarrow{v})-An_{u}(\overrightarrow{r},\overrightarrow{v})
\end{equation}
and%
\begin{equation}
0=\lambda_{l}(v)-\Gamma n_{l}(\overrightarrow{r},\overrightarrow{v})+n_{-}%
(\overrightarrow{r},\overrightarrow{v})BJ_{k}(\overrightarrow{r}%
,\overrightarrow{v})+An_{u}(\overrightarrow{r},\overrightarrow{v})
\end{equation}
Here, the $\lambda_{i}$ are the pump rates which always are assumed to be
proportional to the Maxwellian velocity distributions, $\Gamma$ is the loss
from the states due to all processes other than the maser transition, $\ A$
and $B$ are the Einstein coefficients which are related by $B=c^{2}A/2h\nu
^{3}$ since the statistical weights of the upper and lower states are assumed
to be equal,
\begin{equation}
n_{-}=n_{u}-n_{l}=\frac{\lambda_{u}(v)-\lambda_{l}(v)}{\Gamma}\frac
{1}{1+2BJ_{k}(\overrightarrow{r},\overrightarrow{v})/\Gamma}%
\end{equation}
when an unimportant term in $(A/\Gamma)$ is ignored for simplicity and
\begin{equation}
J_{k}(\overrightarrow{r},\overrightarrow{v})=\int I[\overrightarrow
{r},\widehat{k},\nu=\nu_{0}(1+\widehat{k}\cdot\overrightarrow{v}%
/c)]d\widehat{k}/4\pi
\end{equation}
Here, $I(\overrightarrow{r},\widehat{k},\widetilde{\nu})$ is the intensity of
a ray of maser radiation propagating in the direction\textbf{ }$\widehat{k}%
$\textbf{ }at a location $\overrightarrow{r}$ with frequency $\ \nu$. The
increase in ${I}$ \ when it traverses a differential path length
$ds$\ is given by the radiative transfer equation (EW)
\begin{equation}
dI(\overrightarrow{r},\widehat{k},\nu)/ds=(h\nu_{0}/4\pi)\int
G(\overrightarrow{r},\overrightarrow{v},\widehat{k},\nu)\delta\lbrack\nu
-\nu_{0}(1+\widehat{k}\cdot\overrightarrow{v}/c)]d\overrightarrow{v}%
\end{equation}
where $G(\overrightarrow{r},\overrightarrow{v},\widehat{k},\nu)=n_{-}%
(\overrightarrow{r},\overrightarrow{v})BI(\overrightarrow{r},\widehat{k}%
,\nu)+An_{u}(\overrightarrow{r},\overrightarrow{v})$. It is convenient to
express the foregoing equations in terms of normalized quantities
$\widetilde{I,}$ $\widetilde{J_{k}}$\ ,\ and $\widetilde{\kappa}$\ so that
they become
\begin{equation}
d\widetilde{I}(\overrightarrow{r},\widehat{k},\widetilde{\nu}%
)/d\tau=\widetilde{\kappa}(\overrightarrow{r},\widehat{k},\widetilde{\nu
})\widetilde{I}(\overrightarrow{r},\widehat{k},\widetilde{\nu})+\eta
\exp(-\widetilde{\nu}^{2})
\end{equation}
where $d\tau$ is the differential optical depth parameter and the
normalized maser opacity
\begin{equation}
\widetilde{\kappa}(r,\widehat{k},\widetilde{\nu})=\frac{1}{\pi}\int\frac
{\exp(-w^{2})\delta(\widetilde{\nu}-\widehat{k}\cdot\overrightarrow
{w})}{1+\widetilde{J}_{k}(\overrightarrow{r},\overrightarrow{w})}%
d\overrightarrow{w}%
\end{equation}
is given in terms of the integral over the normalized velocities
$\overrightarrow{w}$ of the masing molecules and
\begin{equation}
\widetilde{J}_{k}(r,w,\theta^{\prime})=\frac{1}{4\pi}\int\sin\theta %
d\theta\int\widetilde{I}(r,\theta,\widetilde{\nu}=w\cos\theta\cos\theta^{\prime}+w\sin\theta\sin\theta^{\prime}\cos
\phi)d\phi
\end{equation}
which expresses the rate of stimulated transitions for molecules with a
normalized velocity of magnitude $w$ in the direction given by the angle
$\theta^{\prime}$ measured relative to the radius vector. In writing these expressions
for $\widetilde{\kappa}$ and $\widetilde{J_{k}}$, we have recognized that the
calculations here will only be applied to geometries (sphere and disk) other
than the linear maser where the intensity only depends on the radial distance
from the center and the angle $\theta$ between this radius and $\widehat{k}.$
The dimensionless $\widetilde{I}(\overrightarrow{r},\widehat{k}%
,\widetilde{\nu})$ and $\widetilde{J}_{k}(\overrightarrow{r},\overrightarrow
{w})$\ in the above equations and throughout the remainder of this Paper are
obtained by dividing the actual intensity by a characteristic ``saturation
intensity''. In the above equations, the saturation intensity is $I_{s}%
=h\nu^{3}_{0}\Gamma/c^{2}A.$ The normalized frequency $\widetilde{\nu}$ is related
to the actual frequency $\nu$ of the radiation by $\widetilde{\nu}%
=(c/v_{D})(\nu-\nu_{0})/\nu_{0}$ where $\nu_{0}$ is the rest frequency of the
transition and $v_{D}$ is the thermal breadth that appears in the Maxwellian
distribution of the velocities, and is reflected in the pumping rates, e.g. $\lambda_{u}(v)=$ $\lambda_{u}(0)\exp[-(v/v_{D})^{2}])$. The normalized
molecular velocity is then $\overrightarrow{w}=\overrightarrow{v}/v_{D}$. In the integrand in equation (8), the frequency $\nu$ also depends upon the anglular difference $\phi$ between the azimuthal angles of $\widehat{k}$ and $\overrightarrow{w}$ measured about $\overrightarrow{r}$ as the axis. The
optical depth parameter $d\tau=\kappa_{0}ds$ \ where $\kappa_{0}=hcv_{D}%
^{2}B\lambda_{-}(0)/4\Gamma$. For simplicity, negligible terms involving the
ratio ($A$/$\Gamma)$ have been omitted (see EW). The contribution of
spontaneous emission in equation (1) is representated by $\eta\exp
(-\widetilde{\nu}^{2})$ where $\eta=(A/\Gamma)[\lambda_{+}(0)/\lambda_{-}(0)]$
and $\lambda_{+}(0),\lambda_{-}(0)$ are the sum and difference of the pump
rates into the upper and lower of the masing states at zero velocity,
respectively. The $\widetilde{J_{k}}$ for the linear and disk masers depend
upon the assumed beaming angles and it is convenient to use other expressions
for the saturation intensities in the calculations for these geometries. The geometry of the calculation is explained in more detail in Appendix B.

\subsection{Linear masers}

For a linear maser, the foregoing equations are greatly simplified and, in the
absence of spontaneous emission, the radiative transfer equation adopts the
familiar form (e.g., Goldreich \& Keeley 1972)
\begin{equation}
d\widetilde{I}(s,\pm\widehat{k}_{0},\widetilde{\nu})/d\tau=\pm\frac
{\widetilde{I}((s,\pm\widehat{k}_{0},\widetilde{\nu})\exp(-\widetilde{\nu}%
^{2})}{1+\widetilde{I}(s,+\widehat{k}_{0},\widetilde{\nu})+\widetilde{I}%
(s,-\widehat{k}_{0},\widetilde{\nu})}+\eta\exp(-\widetilde{\nu}^{2})
\end{equation}
where $\widetilde{I}(s,+\widehat{k}_{0},\widetilde{\nu})$ is the intensity that is
propagating in the direction of increasing $s$ measured along the axis of the maser and $\widetilde{I}(s,-\widehat{k}_{0},\widetilde{\nu})$ propagates in the opposite direction. This dimensionless intensity is obtained by dividing
the actual intensity by the saturation intensity for a linear maser
$I_{s}=4\pi h\nu^{3}_{0}\Gamma/c^{2}A\Delta\Omega$. The meaning of the spontaneous
emission parameter also is modified to incorporate the beaming angle to give
$\eta=(A\Delta\Omega/4\pi\Gamma)[\lambda_{+}(0)/\lambda_{-}(0)]$ in equation
(9). Equation (9) follows from the foregoing general expressions when the
integral over solid angles in equation (4) is performed by treating the
intensities as constant within a solid angle equal to the beaming angle
$\Delta\Omega$ of the maser radiation. This is the standard relationship
between $\widetilde{J}_{k}$ and $\widetilde{I}$ in the linear maser idealization.

\subsection{Spherical masers}

For the spherical maser, one of the integrations for $\widetilde{\kappa
}(r,\widehat{k},\widetilde{\nu})$ can be performed immediately to give%
\begin{equation}
\widetilde{\kappa}(r,\theta,\widetilde{\nu})=\frac{2}{\pi}\int\limits_{0}%
^{\infty}zdz\int\limits_{0}^{\pi}\frac{\exp(-w^{2})}{1+\widetilde{J}%
_{k}(r,w,\theta^{\prime})}d\psi
\end{equation}
where $w^{2}=\widetilde{\nu}^{2}+z^{2}$ and $\ \theta^{\prime}=\arccos
[(z\cos\psi\sin\theta+\widetilde{\nu}\cos\theta)/w]$ (see EW, though note that the variables in EW are not exactly the same as here). Physically, $\psi$ is the angle of projection of $\overrightarrow{w}$ \ on the plane perpendicular to $\overrightarrow{k}$. The remaining integrations are performed most conveniently in these new variables. Equation (10) is then
used together with equations (6) and (8) to perform the numerical computations
for the spherical maser.

\subsection{Disk masers}

To perform the integral in equation (8) over the angle that is perpendicular
to the plane of the thin disk and to relate the result to a physical intensity
which must have a finite angular extent, an idealization is made that is
analogous to that for the linear maser. The intensity is treated as constant
within the beaming angle $\Delta\alpha$ measured perpendicular to the plane of
the disk. In the two dimensional geometry that is appropriate for an
infintesimally thin disk, the integral for $\widetilde{J}_{k}$ then becomes%
\begin{equation}
\widetilde{J}_{k}(r,w,\theta^{\prime})=\frac{1}{2\pi}\int\limits_{0}^{2\pi}%
\widetilde{I}(r,\theta,\widetilde{\nu}=w\cos\theta\cos
\theta^{\prime}+w\sin\theta\sin\theta^{\prime})d\theta
\end{equation}
when the intensity$\ \widetilde{I}\ $is expressed in units of a saturation
intensity that is appropriate for the two dimensional geometry$\ I_{s}=$
$2h\nu^{3}_{0}\Gamma/c^{2}A\Delta\alpha$. Incorporating $\Delta\alpha$ into the
definition of the saturation intensity is analogous to incorporating
$\Delta\Omega$ into the definition of the saturation intensity for the linear
maser. With $\widehat{k}$ restricted to be in the plane of the thin disk, two
of the integrations in equation (7) for $\widetilde{\kappa}(r,\widehat
{k},\widetilde{\nu})$ can be performed immediately to give%
\begin{equation}
\widetilde{\kappa}(r,\theta,\widetilde{\nu})=\frac{1}{\pi}\int\limits_{0}%
^{\infty}dz\exp(-w^{2})[\frac{1}{1+\widetilde{J}_{k}(r,w,\theta_{1}^{\prime}%
)}+\frac{1}{1+\widetilde{J}_{k}(r,w,\theta_{2}^{\prime})}]
\end{equation}
in which $w^{2}=\widetilde{\nu}^{2}+z^{2}$ and $\theta_{1,2}^{\prime}%
=\theta\pm\arccos(\widetilde{\nu}/w)$ to facilitate the final integration. Again, $\theta$ is the angle between the radius vector and the direction of propagation, and $\theta^{\prime}$ is the angle between the radius vector and the molecular velocity. The
computations for disk masers are then performed with these two equations and
with equation (6). The term for spontaneous emission must also be interpreted
as $\eta=$($\Delta \alpha/2)(A/\Gamma)[\lambda_{+}(0)/\lambda_{-}(0)]$ in order to
incorporate the integration over the angle perpendicular to the plane. That
is, equation (6) for the 3D geometry is effectively multiplied by
$\Delta\alpha$/2 to express it in units that are convenient for a disk.

\section{RESULTS OF COMPUTATIONS}

``Saturation'' reflects the importance of maser radiation for altering the
population difference between the masing states, and hence the effect of the
radiation upon $\widetilde{\kappa}.$ In general, this depends upon the
velocity $\overrightarrow{w}$ of the molecule, and upon the direction and
frequency of the radiation. As the benchmark for the degree of saturation of a
maser, we adopt ``R/$\Gamma"$ = $\widetilde{J}_{k}(R_{e},0,0)/2$ for disk and
spherical masers as in EW. This quantity is the rate for stimulated emission
(divided by the loss rate $\Gamma$) for molecules with zero velocity at the
edge of the masing region where the radius is $R_{e}$ . In practice, our
computations for $\widetilde{J}_{k}(R_{e},0,0)/2$ differ negligibly from
$\{\widetilde{\kappa}(R_{e},0,0)^{-1}-1\}/2$ \ which is a possible
alternative, and perhaps more precise, measure of the degree of saturation.
The analogous expression for the degree of saturation at the end of a linear maser of length $L$ is
\ R/$\Gamma=[\widetilde{I}(L,+\widehat{k}_{0},0)+\widetilde{I}(L,-\widehat{k}_{0},0)]/2\simeq\widetilde{I}(L,+\widehat{k}_{0},0)/2$ where $\widehat{k}_{0}$ is in the direction of the emerging radiation. 

The equations are solved by iteration. In each step, a new $\widetilde{J}_{k
}$ is chosen based on the $\widetilde{J}_{k}$ that are obtained in the two
most recent steps. This continues until $\widetilde{J}$ and $\widetilde
{\kappa}$ at all $w$, $\widetilde{\nu}$, and angles in the computational grid
change by less than a specified fractional amount \ between successive
iteration steps. We have, of course, verifed that the results that we present
are insensitive to this fractional difference required for convergence, as
well as to the number of grid points in $w$, $\widetilde{\nu}$, $r$, and in
the angles $\theta$ and $\theta^{\prime}$ at which $\widetilde{J}$ and $\widetilde{\kappa}$ are computed. In the computation
for the disks, $\widetilde{J}$ and $\widetilde{\kappa}$ typically are computed
at 33 values of \ each of the variables $w$, $\widetilde{\nu}$, $r$ and the
angles $\theta$ and $\theta^{\prime}$. For a sphere, we are unable to allow 33 values for all
variables if the computations are to be performed in a ``reasonable'' time of
a few days. Some variable must be constrained to 17 values, but we verify that
the accuracy of the computations is not significantly compromised. These
values (the ``grid'') are uniformly spaced, though the ranges of angles that
are included are reduced to reflect the beaming of the maser radiation. The
ranges of angles depend upon the location in radius at which $\widetilde{J}$
and $\widetilde{\kappa}$ are being computed, as well as upon the degree of
saturation of the model being considered. Additional details about the geometry and about the computations are given in Appendix B and Appendix C.

Two quantities must be specified at the outset of a computation to solve the
foregoing equations for any of the three geometries---

a) the size of the maser expressed in terms of the number of unsaturated
optical depths, that is, $\kappa_{0}R_{e}$ for the disk and spherical
geometries and $\kappa_{0}L$ for the linear maser (where $L$ is the length of
the maser) and

b) the seed radiation parameter---either the external continuum intensity
$\widetilde{I}_{0}$ that is incident onto the outer surface of the masing
region or the spontaneous emission term $\eta$. We find that the results of
computations performed with a specific external intensity $\widetilde{I}_{0}$
and zero spontaneous emission differ negligibly from those performed with zero
external intensity and with the spontaneous emission term $\eta$ set equal to
$\widetilde{I}_{0.}$ In fact, they are indistinguishable except at the very highest
$R/\Gamma$ for which we present the results in the Figures. Hence, we will
give explicit results only for $\eta=0$ and various values of $\widetilde
{I}_{0}$, but with the understanding that the results are applicable when
spontaneous emission is the dominant source for the seed radiation and $\eta$
takes on these same numerical values.

To compare the behavior of the maser properties between the three geometries,
the natural comparison is between computations for which the background
radiation is the same. The background radiation is the same when the rates for stimulated
emission (with $\eta=0$) are equal when no
masing gas is present---that is, when $R/\Gamma$ is the same for the three
geometries in the limit where $\kappa_{0}R_{e}$ and $\kappa_{0}L=0.$ Hence, we
compare computations for which $\widetilde{I}_{0}$ is the same for the sphere
and the disk, and those for a linear maser for which the $I_{0}$ is one-half
of that for the sphere and disk. Once $\widetilde{I}_{0}$ is specified, the
intensity of the radiation that emerges from the maser then depends only upon
$\kappa_{0}R_{e}$ or $\kappa_{0}L$. Instead of designating our results by the
values of\ $\kappa_{0}R_{e}$ and $\kappa_{0}L$, we use the degree of
saturation $R/\Gamma$ which is a measure of the emergent intensity. To
summarize, the computations will be labelled according to their values of
$R/\Gamma$ (which substitutes for $\kappa_{0}R_{e}$ or $\kappa_{0}L$) and
$\widetilde{I}_{0}$.

What are representative values for $\widetilde{I}_{0}$? Consider the 22 GHz
masing transition of the water molecule which is observed and imaged
extensively in a wide variety of astronomical environments. Typical
calculations (e.g., Anderson \& Watson 1993; Wallin \& Watson 1997) for the
pumping rates yield $\lambda_{+}(0)/\lambda_{-}(0)\simeq10-100$. For $\Gamma
$($\simeq1$ s$^{-1}$) which is considered representative for these masers,
then $\eta=(A/\Gamma)[\lambda_{+}(0)/\lambda_{-}(0)]\simeq10^{-7}$ to
$10^{-8}$. For spherical masers, no additional uncertainty enters due to lack
of knowledge about the beaming angle. Hence, $\widetilde{I}_{0}=(A/\Gamma
)(2kT_{b}/h\nu_0)\simeq2T_{b}A/\Gamma\simeq4\times10^{-9}$ $T_{b}$ for isotropic
background radiation with brightness temperature $T_{b}$. The effective value
of $T_{b}$ is expected to be close to the 3K of the cosmic background. For the
linear and disk geometries, these expressions must be multiplied by the
beaming angles --- $\ $by $\Delta\Omega/4\pi$ for the linear geometry and by
$\Delta\alpha/2$ for the disk geometry --- to obtain the analogous estimates
for $\widetilde{I}_{0}$ and $\eta$ for these geometries. Beaming angles for
astrophysical masers are highly uncertain.\ Ratios of ten for the
length-to-width commonly are discussed (for which $\Delta\Omega/4\pi
\simeq10^{-2}$ and $\Delta\alpha/2\simeq10^{-1}),$ though the actual beaming angles 
may easily be smaller by a factor of ten or more. With the understanding that the
spontaneous emission term can be replaced by an $\widetilde{I}_{0}$ of the
same magnitude without changing the results, we can then focus our attention
on computations with $\widetilde{I}_{0}\lesssim10^{-7}$, at least for
applications with the 22 GHz water masers. Analogous estimates suggest that
this range is representative for other strong masers of methanol and 18 cm
hydroxyl, though the range may be extended up to $\sim10^{-5}$ for the circumstellar SiO masers.

\subsection{Spectra}

The flux of radiation at frequency $\widetilde{\nu}$\ \ from a masing sphere
that is seen by a distant observer is
\begin{equation}
\widetilde{F}(\widetilde{\nu},\tau_{0})=2 \Omega_{\mathrm{geom}}\int_{0}^{\pi
/2}\widetilde{I}(R_{e},\theta,\widetilde{\nu})\cos\theta\sin\theta d\theta
\end{equation}
whereas that from an edge-on disk is%
\begin{equation}
\widetilde{F}(\widetilde{\nu},\tau_{0})=\Omega_{\mathrm{geom}}\int_{0}^{\pi
/2}\widetilde{I}(R_{e},\theta,\widetilde{\nu})\cos\theta d\theta%
\end{equation}
and that from a linear maser is simply%
\begin{equation}
\widetilde{F}(\widetilde{\nu},\tau_{0})=\Omega_{\mathrm{geom}}\widetilde
{I}(L,\widehat{k}_{0},\widetilde{\nu})
\end{equation}
in terms of the intensities at the edges which propagate at angles $\theta$ relative to the radius vector (or along $\widehat{k}_{0}$ in the case of linear masers), and where $\Omega_{\mathrm{geom}}$ is the geometrical angular size of the maser.
Note that the flux $\widetilde{F}$ is in units of the saturation intensity
$I_{s}$ for the relevant geometry, and that the unsaturated optical depth parameter $\tau_{0}=2\kappa_{0}R_{e}$ or $\tau_{0}=\kappa_{0}L$ depending upon the geometry.

The behavior of the spectral line breadth of the maser flux as a
function of the degree of saturation is one of the two main results of this
investigation. Representative results are given in Figure 1. There the
spectral line breadth $\widetilde{v}_{1/2}$ --- obtained by solving
\ $\widetilde{F}(\widetilde{v}_{1/2},\tau_{0})=$\ \ $\widetilde{F}(0,\tau
_{0})/2$ --- for the maser flux from disks and spheres is seen to behave in
essentially the same way as that from linear masers. As a function of the
degree of saturation $R/\Gamma$, the breadth first decreases as the maser
becomes more intense but is still unsaturated. Then, when saturation can no
longer be ignored as $R/\Gamma$ approaches one, the profile rebroadens and
ultimately reaches the full thermal Doppler breadth just as occurs for linear
masers. The rebroadening begins slightly later as a function of $R/\Gamma$ for
the disk and the sphere. Once it begins, however, the broadening is somewhat
more rapid than for linear masers. 

Although the maser profiles only approach the Gaussian that has the breadth of
the thermal gas velocities when the maser is saturated, the profiles can be
quite close to Gaussians with smaller breadths in the unsaturated regime. A
useful quantitative measure of the deviation of the maser profile from a
Gaussian is (Watson et al. 2002)
\begin{equation}
\delta=\frac{\int[\widetilde{F}(\widetilde{\nu},\tau_{0})-a_{1}\exp
(-\widetilde{\nu}^{2}/a_{2})]^{2}d\widetilde{\nu}}{2\widetilde{F}_{p}%
^{2}\widetilde{\nu}_{1/2}}%
\end{equation}
where  $\widetilde{F}_{p}=$ $\widetilde{F}(0,\tau_{0})$ is the peak value of
the flux which occurs at the center of the line profile, $2\widetilde{\nu
}_{1/2}$ is the full width at half maximum (FWHM) of this profile, and
($a_{1},a_{2})$ are the parameters obtained in fitting $\widetilde
{F}(\widetilde{\nu},\tau_{0})$ to a Gaussian using the least squares method.
Although $\delta$ can be small in both the saturated and unsaturated regime,
the nature of the deviations from Gaussian in the two regimes is different.
The computed profile is more sharply peaked than a Gaussian (positive
kurtosis) in the unsaturated regime, but is less peaked than a Gaussian
(negative kurtosis) in the saturated regime. The deviation $\delta$ and the
kurtosis K, where 
\begin{equation}
K=\int\widetilde{F}(\widetilde{\nu},\tau_{0})\widetilde{\nu}%
^{4}d\widetilde{\nu}/\sigma^{4}\int\widetilde{F}(\widetilde{\nu},\tau
_{0})d\widetilde{\nu}%
\end{equation}
are shown in Figure 2 for the three geometries. Here, $\sigma$ is the standard
deviation of the distribution $F(\widetilde{\nu},\tau_{0})$. Certain observational
evidence (Watson et al. 2002; Moscadelli, Menten, Walmsley, \& Reid 2003)
indicates that the $\delta$ and $K$ for actual spectral lines are consistent
with the predictions in Figure 2 for spectral lines that have not been rebroadened by the effects of saturation.

As for linear masers, the linebreadths for
the disks and spheres tend to increase with $\widetilde{I}_{0}$ when the maser is not
fully saturated as seen in Figure 3. However, the behavior is relatively
insensitive to $\widetilde{I}_{0}$. In all cases, the line breadths are less
than the thermal breadth and approach the thermal breadth as $R/\Gamma$
exceeds one.

\subsection{Apparent angular sizes}

We calculate the apparent angular size as the region from which a distant
observer sees one-half of the flux to be radiated. The goal is to show how the
apparent angular sizes of the masers depend upon the degree of saturation and
how, if at all, these sizes depend upon the frequency at which they are measured within the spectral line.

At a specific frequency $\widetilde{\nu}$, one-half of the flux\ from a
spherical maser is emitted within an angle $\theta_{1/2}(\widetilde{\nu}%
)$\ \ that is found from equation (13) by solving
\begin{equation}
\int_{0}^{\pi/2}\widetilde{I}(R_{e},\theta,\widetilde{\nu})\cos\theta
\sin\theta d\theta=2\int_{0}^{\theta_{1/2}(\widetilde{\nu})}\widetilde
{I}(R_{e},\theta,\widetilde{\nu})\cos\theta\sin\theta d\theta
\end{equation}
from which it follows that the ratio of the apparent angular size\ to the
geometrical angular size is
\begin{equation}
\Omega_{1/2}^{\widetilde{\nu}}/\Omega_{g}=\sin^{2}\theta_{1/2}(\widetilde{\nu
})
\end{equation}
The apparent angular size based on the total power is found by performing the
additional integration over frequency and finding $\theta_{1/2}$%
\begin{equation}
\int_{0}^{\pi/2}\int_{0}^{\infty}\widetilde{I}(R_{e},\theta,\widetilde{\nu
})\cos\theta\sin\theta d\widetilde{\nu}d\theta=2\int_{0}^{\theta_{1/2}}%
\int_{0}^{\infty}\widetilde{I}(R_{e},\theta,\widetilde{\nu})\cos\theta
\sin\theta d\widetilde{\nu}d\theta
\end{equation}
to give
\begin{equation}
\Omega_{1/2}/\Omega_{g}=\sin^{2}\theta_{1/2}%
\end{equation}
Results of calculations for $\Omega_{1/2}/\Omega_{g}$ as a function of
saturation and for $\Omega_{1/2}^{\widetilde{\nu}}/\Omega_{g}$ as a function
of frequency for representative choices of the saturation are presented in
Figures 4 and 5 for spherical masers. In addition, the detailed variation of the intensity $\widetilde{I}(R_{e}%
,\theta,\widetilde{\nu})$ with $\theta$ and $\widetilde{\nu}$ from a representative calculation for a saturated
spherical maser is shown in Appendix A.

The analogous apparent sizes are given for disks viewed edge-on in Figures 4
and 6. Only the angular extent along the plane of the disk varies since the
disks here are idealized as having zero (or very small and constant) thickness. For disks, $\theta_{1/2}(\widetilde{\nu})$ is found, using equation (14), from%
\begin{equation}
\int_{0}^{\pi/2}\widetilde{I}(R_{e},\theta,\widetilde{\nu})\cos\theta d\theta=2\int_{0}^{\theta_{1/2}(\widetilde{\nu})}\widetilde{I}(R_{e}%
,\theta,\widetilde{\nu})\cos\theta d\theta
\end{equation}
and then the apparent to geometrical angular extent along the plane of the
disk at frequency $\widetilde{\nu}$ is%
\begin{equation}
\Omega_{1/2}^{\widetilde{\nu}}/\Omega_{g}=\sin\theta_{1/2}(\widetilde{\nu})
\end{equation}
Similarly, the apparent angular sizes for edge-on, masing disks based on\ the
total power are found from%
\begin{equation}
\int_{0}^{\pi/2}\int_{0}^{\infty}\widetilde{I}(R_{e},\theta,\widetilde{\nu
})\cos\theta d\widetilde{\nu}d\theta=2\int_{0}^{\theta_{1/2}}\int_{0}^{\infty
}\widetilde{I}(R_{e},\theta,\widetilde{\nu})\cos\theta d\widetilde{\nu}d\theta
\end{equation}
and
\begin{equation}
\Omega_{1/2}/\Omega_{g}=\sin\theta_{1/2}%
\end{equation}

\noindent
In Figure 4, we also make comparisons with the results from simpler, approximate calculations that are appropriate in limited regimes of saturation (see Goldreich \&\ Keeley 1972). In the unsaturated limit, the intensity has the familiar exponential dependence on path length. When this form for the intensity is inserted into equations (18) and (22), the apparent angular sizes can be obtained in a nearly analytic calculation. In the partially saturated regime, the reasoning is more involved but still leads to a result that can be evaluated nearly analytically.

In Figures 5 and 6, the normalized spectral line profile of the flux is shown to demonstrate that apparent angular sizes are relatively insensitive to frequency at frequencies where the flux is appreciable.  Significant variations in the sizes as a function of frequency do occur.  However, they are only in the far wings of the line profile.
 
\section{COMPARISON WITH MASING \\
SPHERES AND DISKS NEAR A STRONG \\
CONTINUUM SOURCE}

The masers in the foregoing Sections 2 and 3 are isolated. Hence, the seed
radiation is either the external, isotropic radiation background or the result
of spontaneous emission within the maser. For some masers, it is likely that a
strong continuum source is nearby and dominates as the source of the seed
radiation. To obtain an indication about how the behavior of the spectral line
breadths and angular sizes might be different from that of the isolated
masers, we consider the idealized geometry where the radiation that is
incident on the maser from the continuum source can be approximated as
parallel rays. This idealization would be strictly applicable when the
continuum source is small and far from the masing gas. The rays of maser
radiation then remain parallel within the masers and after they emerge from
the masing region. As a result, there is no interaction between the rays and
the maser radiation that emerges along a ray is the same as that from a linear
maser of the same length. The total flux from the sphere or disk is then just
the sum of the radiation from a series of linear masers with appropriately
varying lengths. That is, for the sphere%
\begin{equation}
\widetilde{F}(\widetilde{\nu},\tau_{0})=2\pi\int_{0}^{R_{e}}\widetilde
{I}[L(b),\widehat{k}_{0},\widetilde{\nu}]bdb
\end{equation}
where $\widetilde{I}[L(b),\widehat{k}_{0},\widetilde{\nu}]$ is the solution to
\begin{equation}
d\widetilde{I}(s,\widehat{k}_{0},\widetilde{\nu})/d\tau=\frac{\widetilde{I}((s
,\widehat{k}_{0},\widetilde{\nu})\exp(-\widetilde{\nu}^{2})}{1+\widetilde{I}(s
,\widehat{k}_{0},\widetilde{\nu})}%
\end{equation}
for a linear maser of length $L$ that is analogous to equation (9) but is appropriate when
the maser radiation propagates in only the one direction because of the strong continuum source.
Here, $L(b)$ is the path length for a ray that passes
through the sphere at impact parameter $b$, and $\tau_{0}=2\widetilde{\kappa}_{0}R_{e}$ is the unsaturated optical depth parametr for a ray that passes through the center of the sphere.For the disk (with the incident rays
parallel to the plane of the disk),%
\begin{equation}
\widetilde{F}(\widetilde{\nu},\tau_{0})=2\int_{0}^{R_{e}}\widetilde
{I}[L(b),\widehat{k}_{0},\widetilde{\nu}]db
\end{equation}
With these expressions for $\widetilde{F}$, the spectral line breadth
$\widetilde{\nu}_{1/2}$ and the apparent angular sizes $\theta_{1/2}$ and
$\theta_{1/2}(\widetilde{\nu})$ can be obtained in exactly the same way as was
done in Section 3 for the $\widetilde{F}$ that result from isotropic seed
radiation. These are shown in Figures 7, 8 and 9. To provide a direct
comparison with the linebreadths and the angular sizes for isotropic seed
radiation in Figures 1 and 4, the linebreadths in Figure 7 and the angular
sizes in Figure 8 are computed for $\widetilde{I}_{0}=10^{-9}$. However, the
apparent angular sizes as a function of frequency in Figure 9 are obtained for
the larger $\widetilde{I}_{0}=10^{-6}$ which probably is more realistic when a
strong continuum source is dominant. As for the case of isotropic seed
radiation, the spectral line profile in Figure 7 rebroadens in the
neighborhood of $R/\Gamma\simeq1$ to the full thermal Doppler breadth. Similarly,
the variation of the apparent angular size with frequency in Figure 9 is
unimportant at frequencies within the spectral line profile for which the flux
is appreciable. The variation of $\ \theta_{1/2}(\widetilde{\nu})$ with
frequency is not shown for a sphere. It is similar to that of the disk. In
contrast, the variation in Figure 8 of the apparent angular size with
$R/\Gamma$ is quite different from that for isotropic seed radiation. The
limits for large and small $R/\Gamma$ can readily be obtained. In the fully
saturated\ regime (large $R/\Gamma)$, $\widetilde{I}[L(b),\widehat{k}_{0},\widetilde
{\nu}]$ is just proportional to the length of the ray with impact paramenter
$b$. This yields $\Omega_{1/2}/\Omega_{g}=0.37$ for a sphere and $\Omega
_{1/2}/\Omega_{g}=0.40$ for a disk,  in good agreement with the angular sizes
in Figure 8 at log$R/\Gamma=3$. In the unsaturated limit, $\widetilde{I}%
[L(b),\widehat{k}_{0},\widetilde{\nu}]$ is proportional to
exp(constant$\times$length). For log$R/\Gamma=-2$, this gives $\Omega
_{1/2}/\Omega_{g}=0.076$ and $0.16$, respectively, for the sphere and disk --- again in good agreement the results in Figure 8.  Note that for $R/\Gamma\ll1$, seed radiation due to an isotropic flux and to illumination from one side yield nearly equal  $\Omega
_{1/2}/\Omega_{g}$. That they should be nearly equal can be seen by inserting the expression for the intensity---with its simple, exponential dependence on path length for the unsaturated limit---into
equations (18) and (22).

\section{DISCUSSION}

The foregoing calculations have been performed for the standard idealization
of an astrophysical maser---two energy levels and rate equations with
``phenomenological'' pumping $\Lambda$ and decay $\Gamma$ rates for the
molecular populations as used in equations (1) and (2). This standard
idealization has been the basis for all previous investigations of the
spectral line profiles and angular sizes of astrophysical masers that consider
deviations from the linear maser approximation. The basic equations (6)-(8) that form the starting point are the same as those of other investigators (e.g., Bettwieser 1974; Neufeld 1992) as well as for EW. The main results are the
demonstrations summarized in Figures 1 and 4. (a) The spectral line rebroadens
to the full Doppler width in the saturated regime for disk and spherical
masers in essentially the same way as for linear masers and (b) the apparent
angular sizes of these masers vary only slightly with frequency when the sizes
are measured at frequencies where the flux is significant ( $\gtrsim$ 0.1 of
the flux at line center). While these conclusions were reached in the previous
investigation (EW), the solutions were obtained there by using approximations to
the full equations. That these approximations are of sufficient accuracy seems
to have been unclear to some readers. In contrast, the calculations here solve
the full equations that describe the rate equations and radiative transfer for
the standard idealization for astrophysical masers. In that sense, the
solutions are ``exact''. Though the calculations do depend, in detail, on the
intensity of the seed radiation---isotropic background radiation, a nearby
continuum source, or spontaneous emission, the conclusions (a) and (b) above
are independent of the type of seed radiation and of its intensity---at least within a plausible range of values for the intensity. It should
not be surprising that the spectral line rebroadens with increasing degree of
saturation in a way that is relatively independent of the geometry. The maser
radiation from a disk or a spherical maser becomes highly beamed when the
maser is saturated. Highly beamed rays are nearly parallel, and are thus
similar to those of linear masers where the rays are exactly parallel.

All investigations to date of the above issues about the spectral line
profiles and apparent angular sizes of masers with finite cross sections are
based on the same idealized description as is used in our calculations. Actual
masing regions in astronomy probably deviate at least to some degree from
these idealized conditions. Velocity gradients, as well as variations in the
composition and excitation may be present.\ The observed maser radiation may
be a result of amplification by spatially separated components. The masing transition may consist of hyperfine components (22 GHz water). Finally, $\Lambda$ and
$\Gamma$ are approximations to the results obtained by solving a set of rate
equations for the velocity-dependent populations of the non-masing, as well as
the masing, excitation states of the molecule. The rate equations ordinarily
involve non-masing radiative transitions (usually infrared) and collisional
excitations. These processes may also cause velocity relaxation without a
change in the excitation state of the molecule. In view of these
complications, the reduction of the rate equations to two-levels and
phenomenological ($\Lambda,\Gamma)$ parameters provides a remarkably good
description when it has been tested (Anderson \&\ Watson 1993). Calculations
based on the standard idealization thus provide benchmark results. The
possibility cannot be excluded that some of the additional complexities that
may occur in actual astrophysical masers---which neither we nor others have investigated---will lead to deviations
from these benchmark results. It seems unlikely, however, that the basic
conclusion that rebroadening of a spectral line occurs until the full Doppler
breadth is reached will be altered.

The observed spectral line breadths of maser radiation often are narrower than the thermal Doppler breadths for the gas temperatures that are believed to be appropriate for the masing gas based on likely  pumping schemes or other considerations. Investigators are reluctant to accept the interpretation that these masers are unsaturated because their fluxes are relatively constant in time, because the inferred luminosities seem to require a high efficiency for the pumping, etc.
The natural resolution is that the rate $R$ for stimulated emission of these masers satisfies $\Gamma<R<$ the rate $\gamma$ for velocity relaxation, as recognized long ago by Goldreich \&\ Kwan (1974). 
In this regime, the line breadth remains narrow as for unsaturated masers, but the gain (linear with the length of the maser) and the efficiency of the pumping are the same as for a saturated maser.
The behavior of the line breadth as a function of the degree of saturation has been calculated with velocity relaxation for a linear maser (Nedoluha \&\ Watson 1991; also Anderson \&\ Watson 1993) and explicitly shown to be the same as without velocity relaxation if $\gamma$ is substituted for $\Gamma$. Although velocity relaxation has not been incorporated into the calculations of this Paper, its effect on the line breadth is expected to be similar to that for the linear maser.

We are grateful to A. Ostrovskii and A. Sobolev for help in finding
the best iteration algorithms to use. We also acknowledge support from NSF
Grant AST99-88104.

\appendix

\section{Examples of Intermediate Results}
In Figures 10,11, and 12, we show examples of $\kappa(R_{e},\theta ,\widetilde{\nu})$, $\widetilde{J}_{k}(R_{e},w,\theta^{\prime})$ and $I(R_{e}%
,\theta,\widetilde{\nu})$ that are computed at the surface of a masing sphere that is radiatively saturated. In the calculations, quadratic
interpolation is used for $\widetilde{\kappa}$ and $\widetilde{J}_{k}$
near the symmetry points where $\theta$ or $\theta^{\prime}$ is equal to $0$ or $90$ degrees, and at $\widetilde{\nu}$ and $w=0$.

\section{The Geometry for the Calculation of Isolated Disk and Spherical Masers}

To perform the computation with a disk or spherical maser for the intensity $I(r,\theta ,\nu )$ at radius $\overrightarrow{r}$ propagating in the direction $\widehat{k}$ (which makes an angle $\theta $ with $\overrightarrow{r}$) and with frequency $\nu $, equation (6) is integrated along the paths of the rays such as the ray shown in Figure 13. For example, at the integration step that is a distance $s$ along this path with impact parameter $b$ shown in Figure 13, $\widetilde{\kappa }$ \ from the previous iteration is evaluated at the location $\overrightarrow{r}^{\prime \prime }$ for the angle $\theta ^{\prime \prime }$ to compute the right hand side of equation (6). After the intensities at $\overrightarrow{r}$ for rays at a sufficient number of angles have been computed, the integral in equation (8) is performed to find $\widetilde{J_{k}}$ at the location $\overrightarrow{r}$ for the molecular velocity of magnitude $w$ that is in the direction $\theta ^{\prime }$ shown in Figure 13. In turn, these values of $\widetilde{J_{k}}(r,w,\theta ^{\prime })$ are inserted into equation (10) for a spherical maser or equation (12) for a disk maser to obtain a new value for $\widetilde{\kappa }(r,\theta ,\nu ).$ In the computation for a spherical maser, $\overrightarrow{w}$ ordinarily is not in the same plane as $\overrightarrow{r}$ and $\widehat{k}$. The $\overrightarrow{r}\widehat{k}$ plane differs from the $\overrightarrow{r}\overrightarrow{w}$ plane by the angle of rotation $\phi $ about $\overrightarrow{r}$ as the axis of rotation.

\section{Additional Information on the Numerical Methods}

For the reader who might be interested in repeating our numerical
calculations, here are a few details. In order to place more integration
points near the center of the sphere (or disk), where as discussed by, e.g.
Goldreich and Keeley (1972), there is a transition region between saturated
and unsaturated, we distributed the points at radii proportional to $r^{2}$.
Recognizing that the radiation is beamed, we distributed the angular points
uniformly over the range $0\leq\theta\leq\theta_{\max}$ with $\theta_{\max
}=180^{\mathrm{o}}/(1+0.2\kappa_{0}r)$. Thus, $\theta$ covers the entire angular
range at the center, and decreases toward the perimeter and with saturation
since $\kappa_{0}r$ increases with increasing saturation. The coefficient 0.2
in the expression for $\theta_{\max}$ was chosen by trial and error. We also
experimented with other forms for $\theta_{\max}$ and for the $r$ dependence.
We examined the choices which we adopted for the distributions of points in
angle and in radius by plotting various distributions of the computed
quantities after the fact.

The equations were solved by cross iteration between the equation for
$\widetilde{J}_{k}$ and the equation for $\widetilde{\kappa}$ using the linear
superposition $\widetilde{J}_{k}=f\widetilde{J}_{k}^{1}+(1-f)\widetilde{J}%
_{k}^{2}$ to obtain $\widetilde{J}_{k}$ for the next iteration step from the
results of the most recent iteration step ($\widetilde{J}_{k}^{1}$) and the
next most recent iteration ($\widetilde{J}_{k}^{2}$) step. We used $f=0.4$ for the
first 15 iterations, $f=0.2$ for the next 15, and $f=0.1$ for the subsequent
iterations which typically number about 100 for high saturation (with fewer interations for lower saturations) to achieve the convergence criterion that the maximum change between iteration steps in $\widetilde{J}_{k}$ and $\widetilde{\kappa}$ at any integration point be less than one
percent. Further tightening of the convergence criterion led to negligible
change in the results. Computations are performed first for a completely
unsaturated case (small $\kappa_{0}R_{e}$)\ where $\widetilde{J}_{k}%
=$constant is taken as the initial value for the iterations. For subsequent, larger values of $\kappa_{0}R_{e}$ (where saturation may be important), the
converged $\widetilde{J}_{k}$ from the computation for the previous choice
for $\kappa_{0}R_{e}$ are taken as the initial values after rescaling in $r$. 

The calculation did not require a supercomputer, even for the case of the
spherical geometry which is the most difficult. All computations were
performed on a PC with an AMD XP1900 microprocessor running under MS Windows.
When the code is compiled using Compaq Fortran, the numerical computations for
single curve in Figures such as Figure 1 requires about 36 hours for a sphere.
When the code is compiled by Microsoft Visual Studio.Net optimizing C++, the
analogous running times were typically 30\% less.

\clearpage

\begin{figure}
\epsscale{0.8}
\includegraphics[angle=270,width=1.0\textwidth]{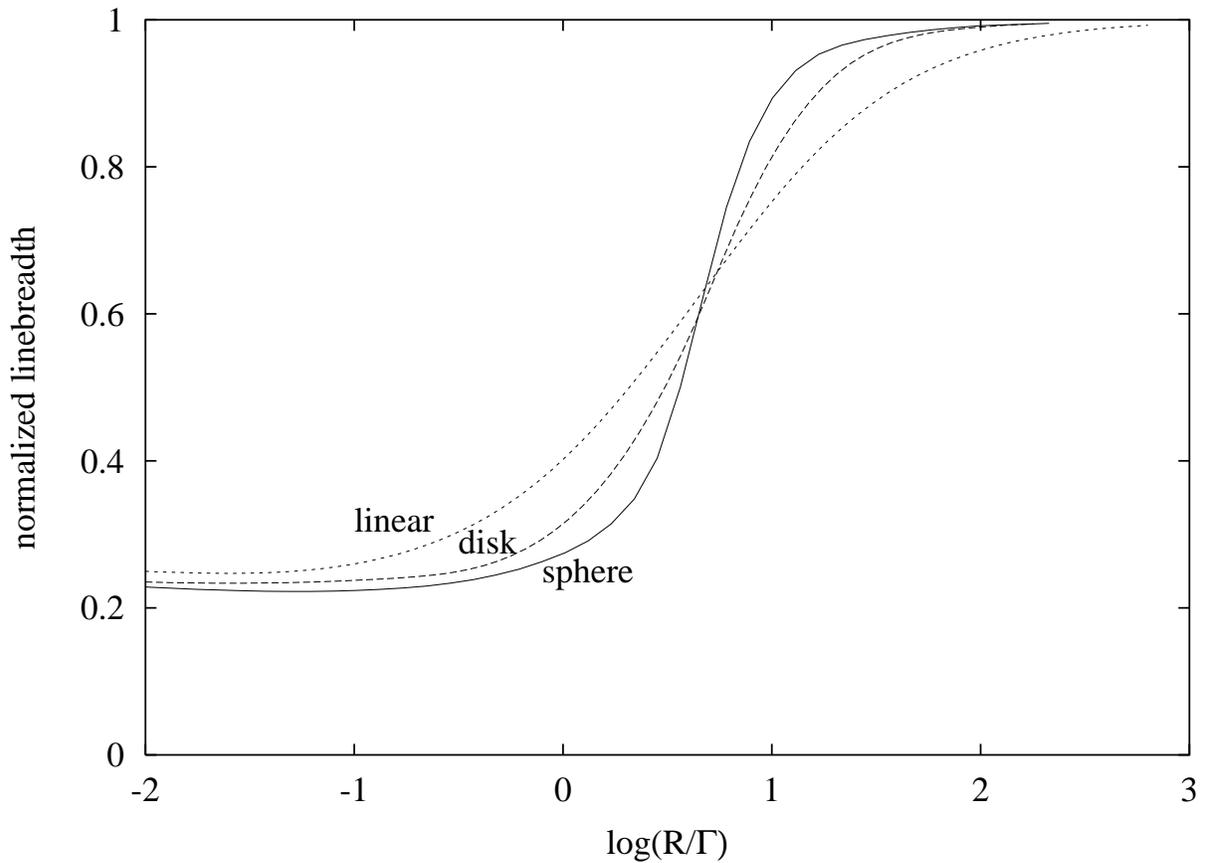}
\caption{Comparison of the spectral line breadth (FWHM) divided by the
thermal Doppler breadth (FWHM) for the flux of observed maser radiation from
linear, thin (edge-on) disk and spherical masers as a function of the log of the degree
of saturation $R/\Gamma$\ (with $\widetilde{I}_{0}=10^{-9}$).}
\end{figure}

\begin{figure}
\epsscale{0.8}
\includegraphics[angle=270,width=1.0\textwidth]{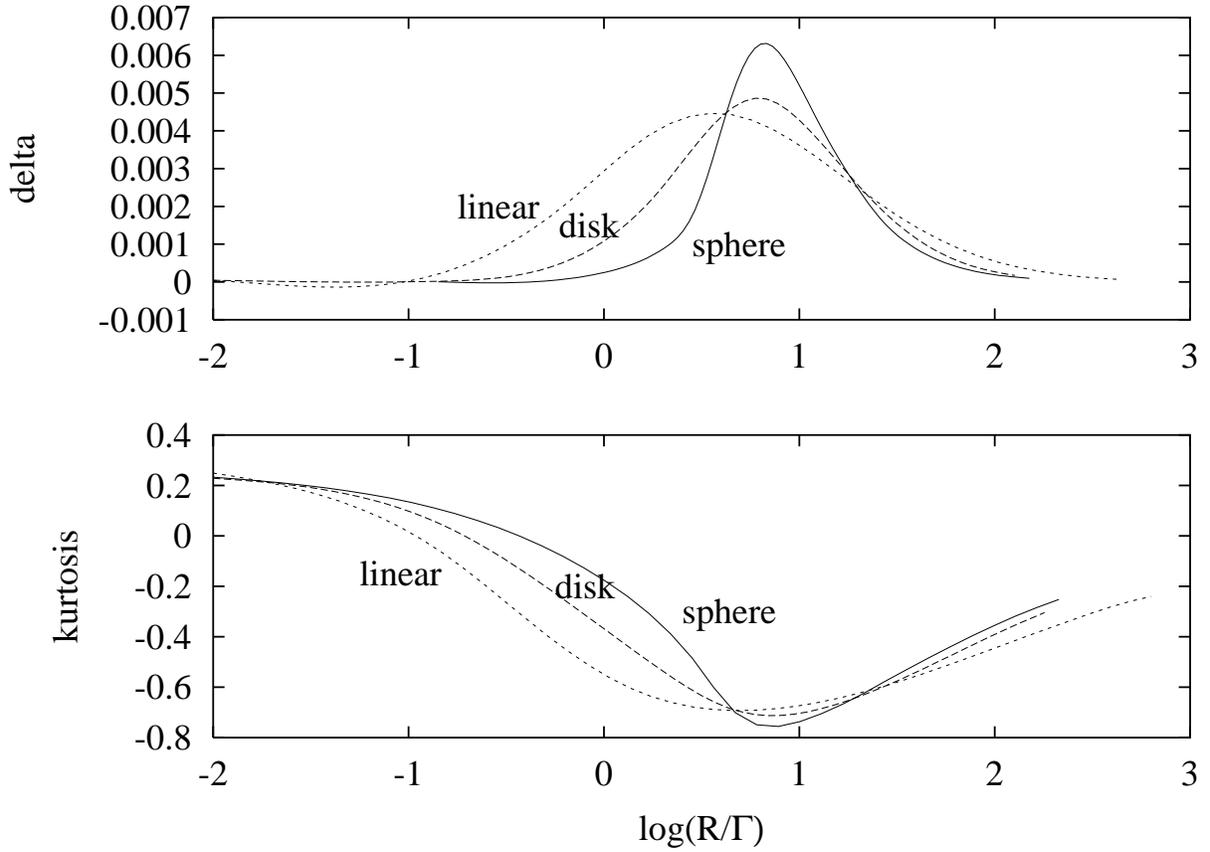}
\caption{(upper panel) The quantity $\delta$ which measures the deviation of
the profile of the radiation from the Gaussian that best fits the
profile, as a function of the log of the degree of saturation, for the linear,
thin (edge-on) disk and spherical masers (with $\widetilde{I}_{0}=10^{-9}$). (lower panel) The
kurtosis K for the same profiles.}
\end{figure}

\begin{figure}
\epsscale{0.8}
\includegraphics[angle=270,width=0.6\textwidth]{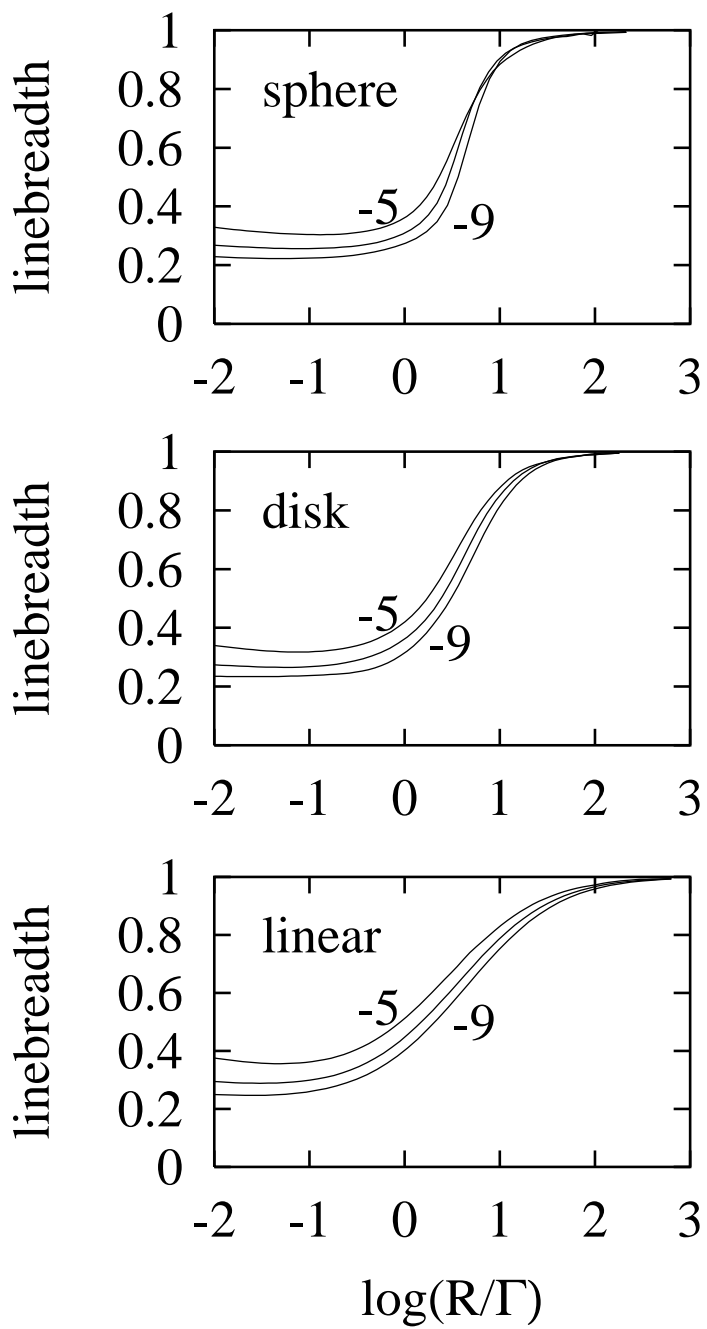}
\caption{The dependence of the spectral linebreadth of the observed maser
radiation on $\widetilde{I}_{0}$ for $\widetilde{I}_{0}=10^{-5},10^{-7},$and $10^{-9}$.\ \ The
spectral line breadth (FWHM) divided by the thermal Doppler breadth (FWHM) for
the flux of observed maser radiation from linear, thin disk, and spherical
masers is given a function of the log of the degree of saturation.}
\end{figure}

\begin{figure}
\epsscale{1.0}
\includegraphics[angle=270,width=1.0\textwidth]{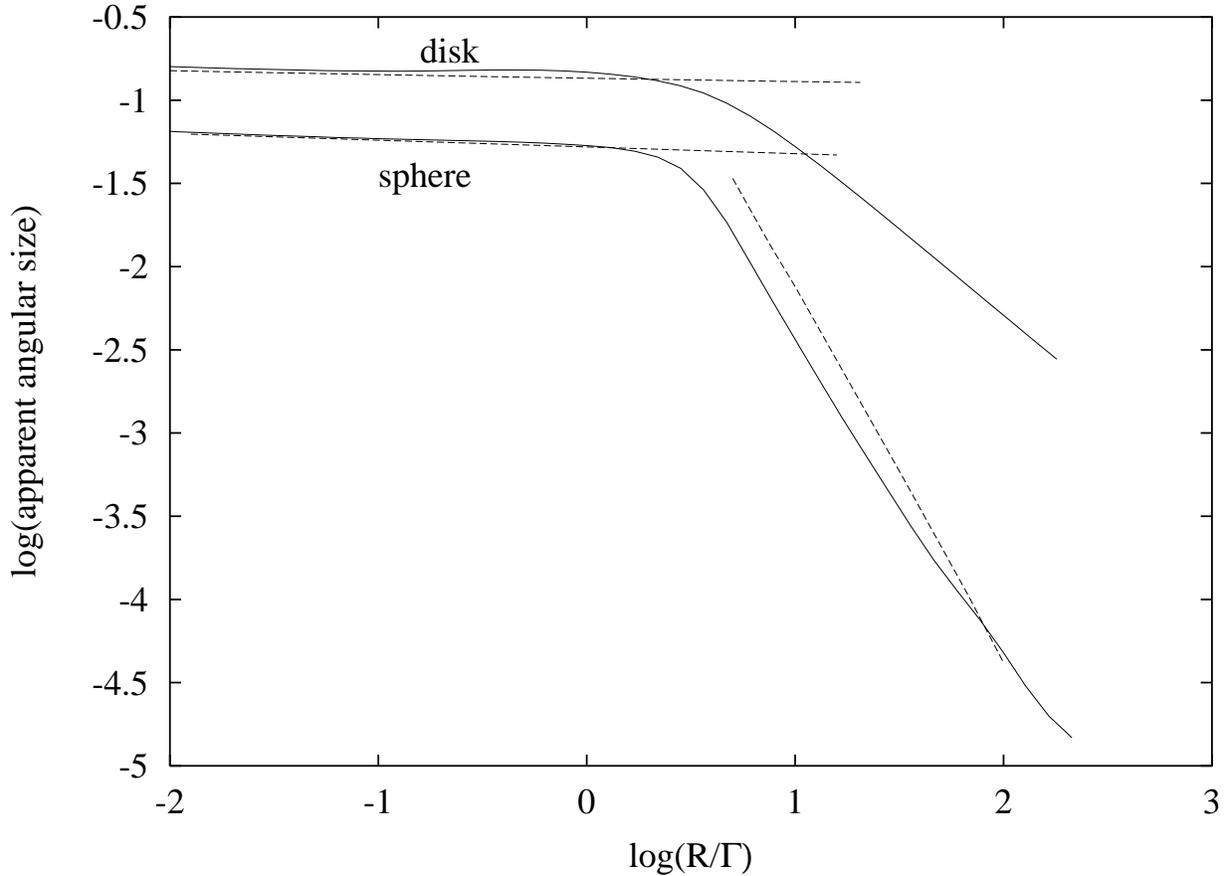}
\caption{The log of the ratio $\Omega_{1/2}/\Omega_{g}$ of the apparent
angular size to the geometrical angular size for thin (edge-on) disk and spherical
masers as measured by the total power, as a function of the log of the degree
of saturation (with $\widetilde{I}_{0}=10^{-9}$). Apparent angular sizes based on approximations that are appropriate in the
unsaturated and partially saturated regimes are given by the straight lines (see text).}
\end{figure}

\begin{figure}
\epsscale{0.8}
\includegraphics[angle=270,width=1.0\textwidth]{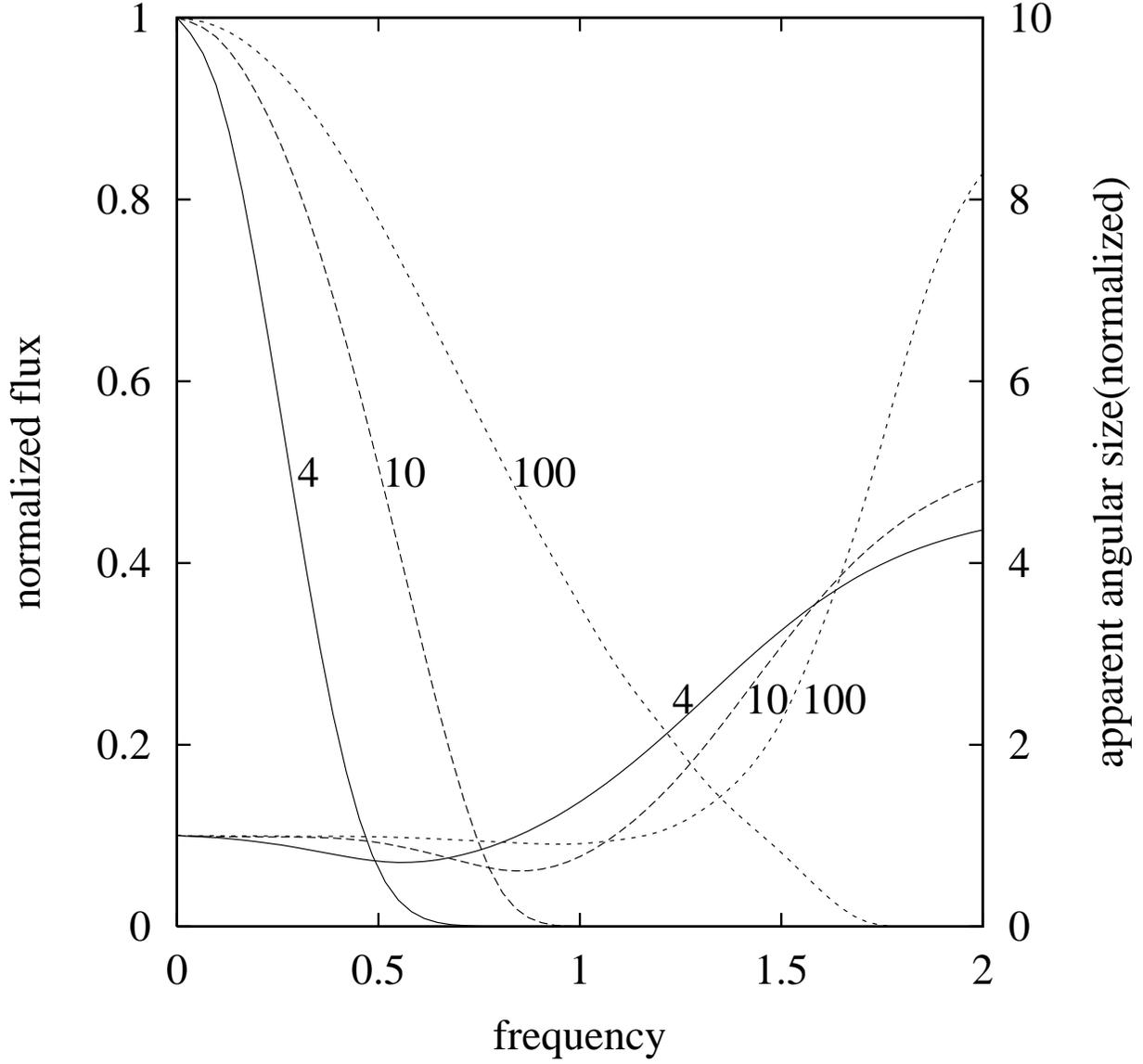}
\caption{The ratio $\Omega_{1/2}^{\widetilde{\nu}}/\Omega_{g}$ of the
apparent angular size to the geometrical angular size for a spherical maser as
measured by the flux at frequency $\widetilde{\nu}$. The ratio is shown for values of the degree of saturation $R/\Gamma$ ($=4,10,100$) as
indicated (with $\widetilde{I}_{0}=$ $10^{-9}$). The normalized flux (normalized by the flux at
$\widetilde{\nu}=0$) also is shown as a function of frequency at the same
values of $R/\Gamma$.}
\end{figure}

\begin{figure}
\epsscale{0.8}
\includegraphics[angle=270,width=1.0\textwidth]{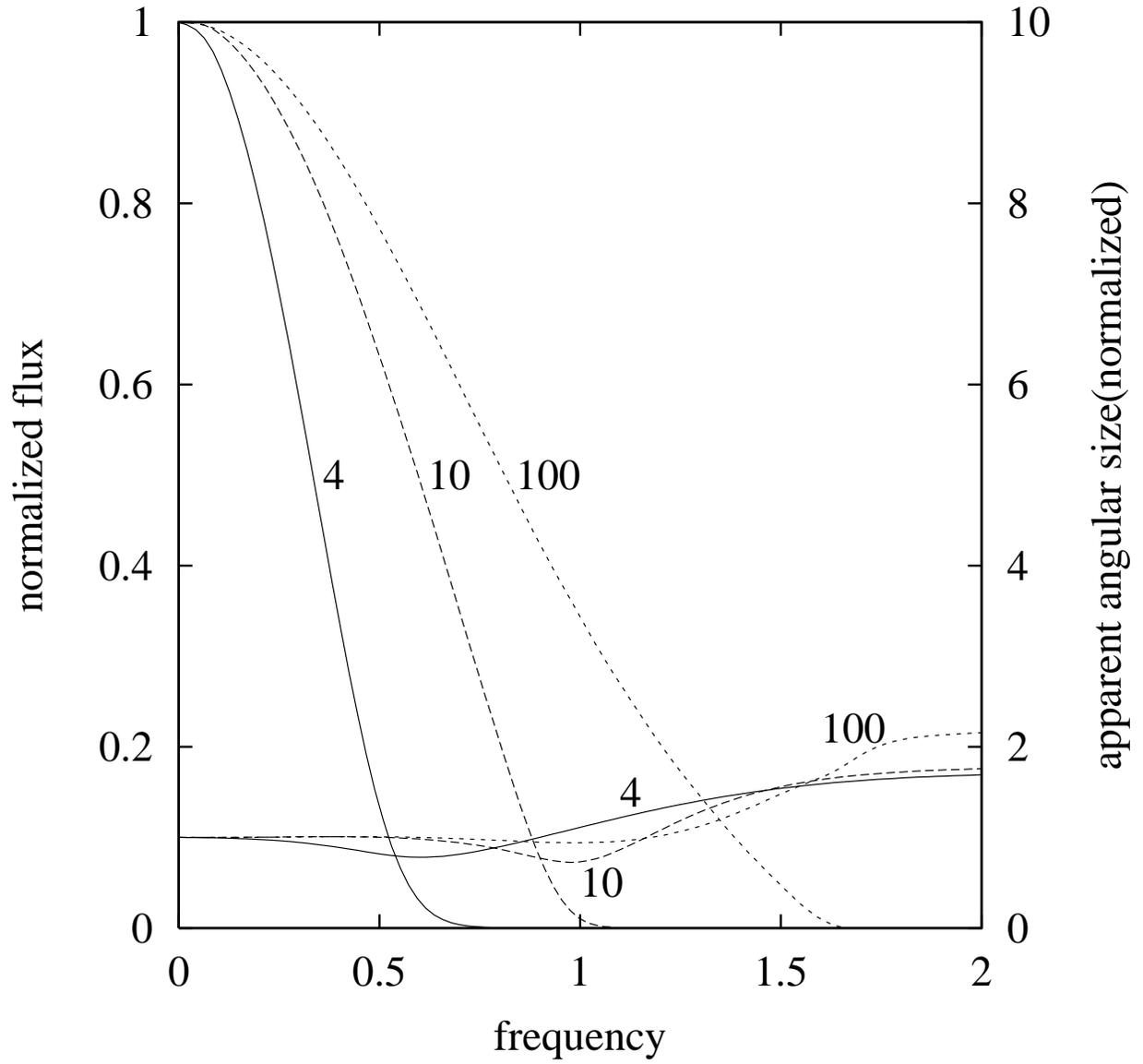}
\caption{Similar to Figure 5 except the maser is in a thin disk.}
\end{figure}

\begin{figure}
\epsscale{0.8}
\includegraphics[angle=270,width=1.0\textwidth]{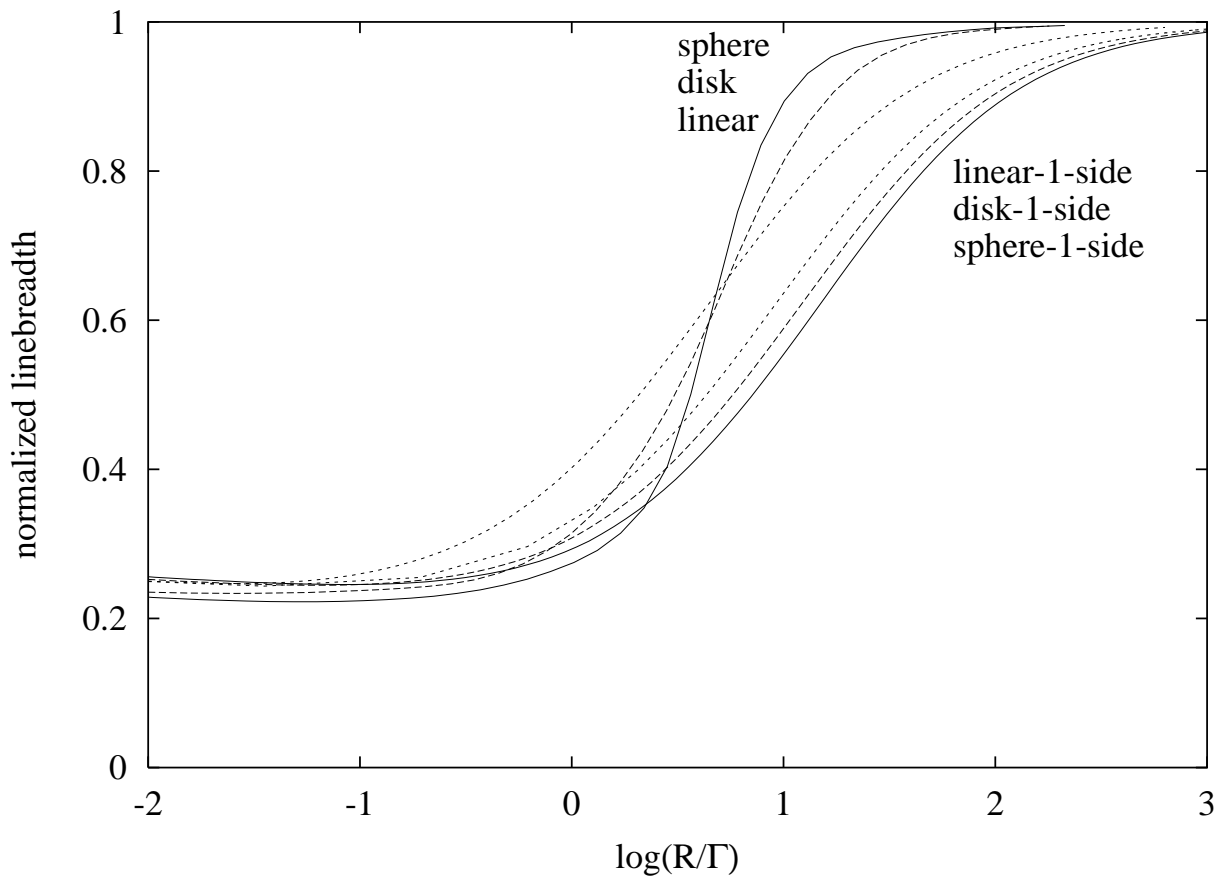}
\caption{Same as Figure 1 except that a strong, external continuum source on
one side (the far side) provides the seed radiation for these masers. The spectral line
breadth (FWHM) divided by the thermal Doppler breadth (FWHM) is shown for
linear (dots), thin disk (dashes), and spherical (solid line) masers. For comparison purposes, the spectral
line breadths in Figure 1 for an isotropic background are shown as well. The
background continuum intensity for all of the computations in this Figure is
$\widetilde{I}_{0}=10^{-9}$.}
\end{figure}

\begin{figure}
\epsscale{1.0}
\includegraphics[angle=270,width=1.0\textwidth]{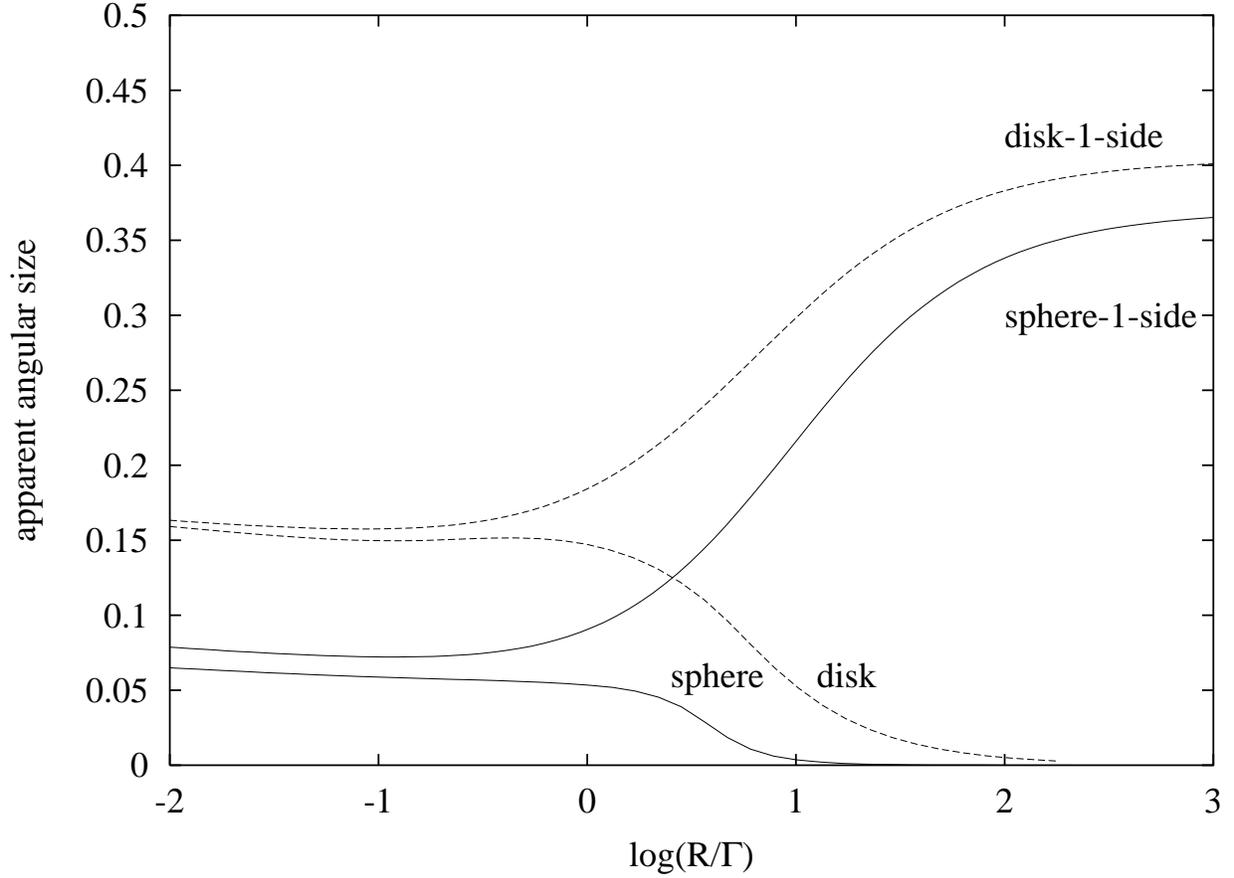}
\caption{Similar to Figure 4 except that a strong, external continuum source
on one (the far side) provides the seed radiation for these masers. The 
ratio $\Omega_{1/2}/\Omega_{g}$ of the apparent angular size to the
geometrical angular size for the thin disk and spherical maser as measured by
the total power is shown. The ratio is given as a function of the log of the degree of
saturation $R/\Gamma$. For comparison purposes, the analogous ratios
$\Omega_{1/2}/\Omega_{g}$ from Figure 4 for thin disk and spherical masers
with an isotropic background are shown as well. The background continuum
intensity for all of the computations in this Figure is $\widetilde{I}_{0}=10^{-9}$.}
\end{figure}

\begin{figure}
\epsscale{0.8}
\includegraphics[angle=270,width=1.0\textwidth]{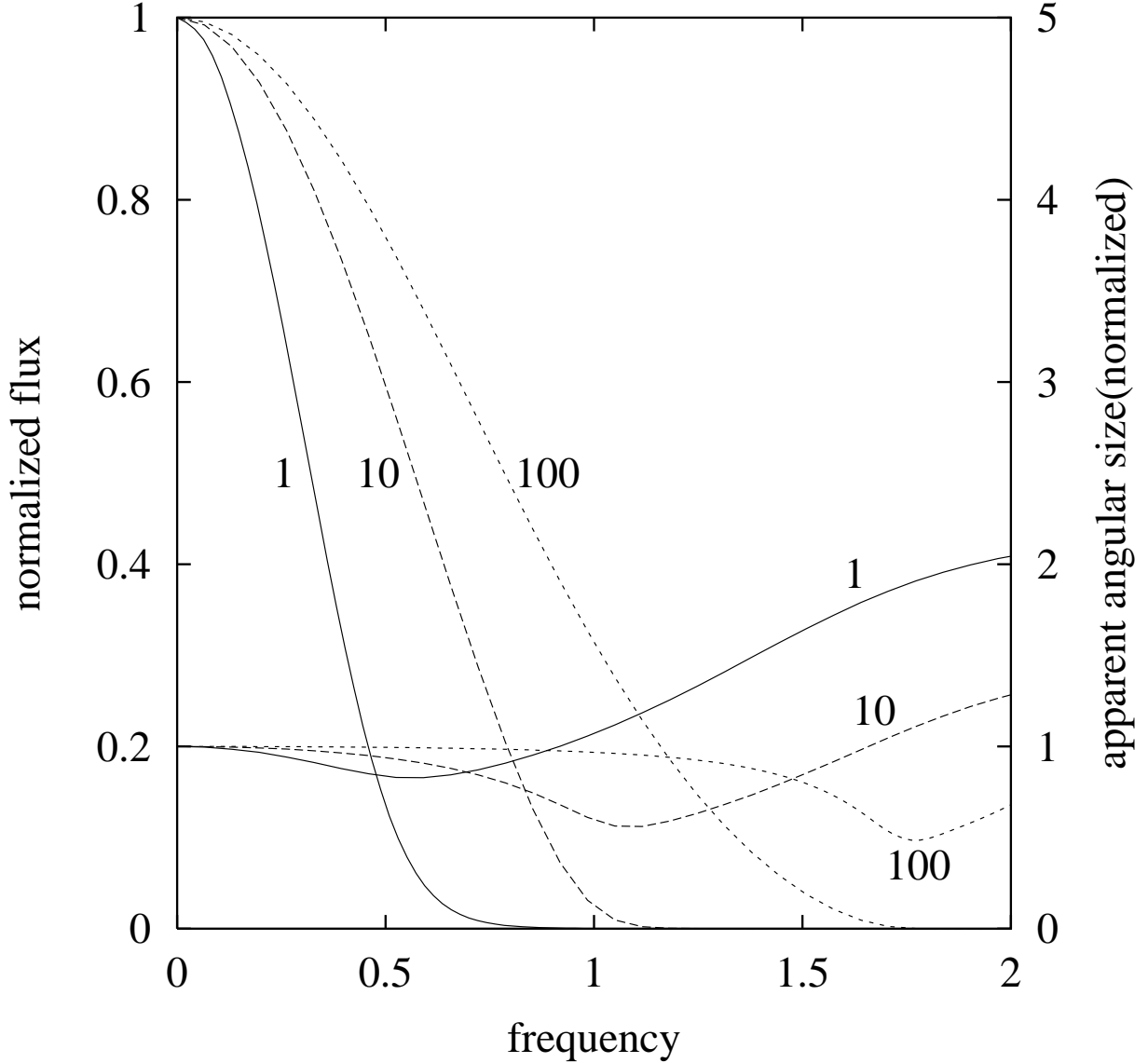}
\caption{Similar to Figure 5 except that a strong, external continuum source
on the far side provides the seed radiation for these masers. The ratio
$\Omega_{1/2}^{\widetilde{\nu}}/\Omega_{g}$ of the apparent angular size to
the geometrical angular size is shown for a thin disk maser as measured by the flux at
frequency $\widetilde{\nu}$. The ratio is given for several values of the
degree of saturation $R/\Gamma$ ($=1,10,100$) as indicated (with $\widetilde{I}_{0}=10^{-6}$). The normalized flux (normalized by the flux at $\widetilde{\nu}=0$)
also is shown as a function of frequency at the same values of $R/\Gamma$.}
\end{figure}

\begin{figure}
\epsscale{0.8}
\includegraphics[angle=270,width=1.0\textwidth]{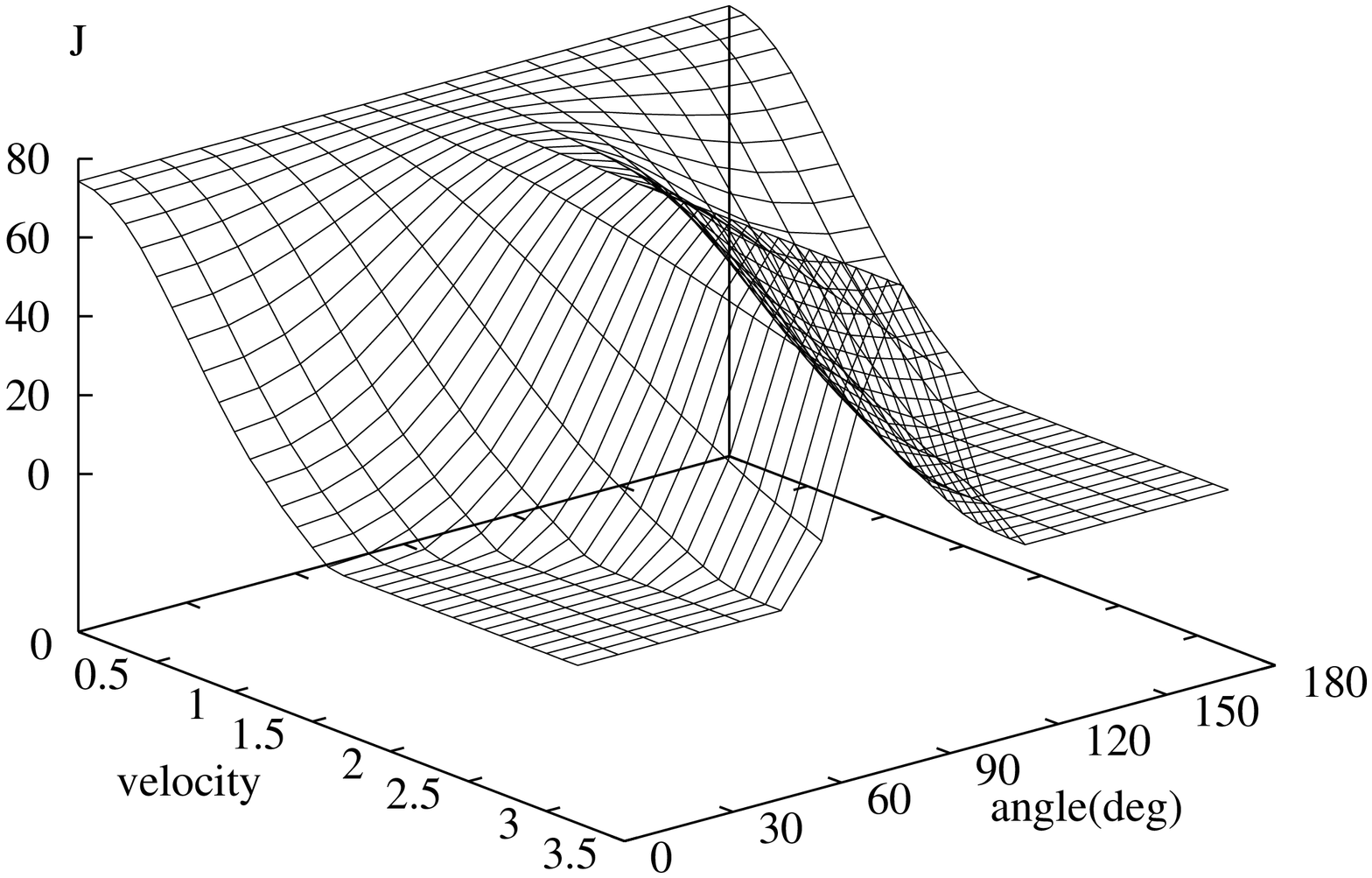}
\caption{An example of the function $\widetilde{J}_{k}(R_{e},w,\theta^{\prime})$ in the
computations for a spherical maser with seed radiation due to an isotropic
background ($\widetilde{I}_{0}=10^{-9}$). In the computations for this example,
$R/\Gamma=52.$}
\end{figure}

\begin{figure}
\epsscale{0.8}
\includegraphics[angle=270,width=1.0\textwidth]{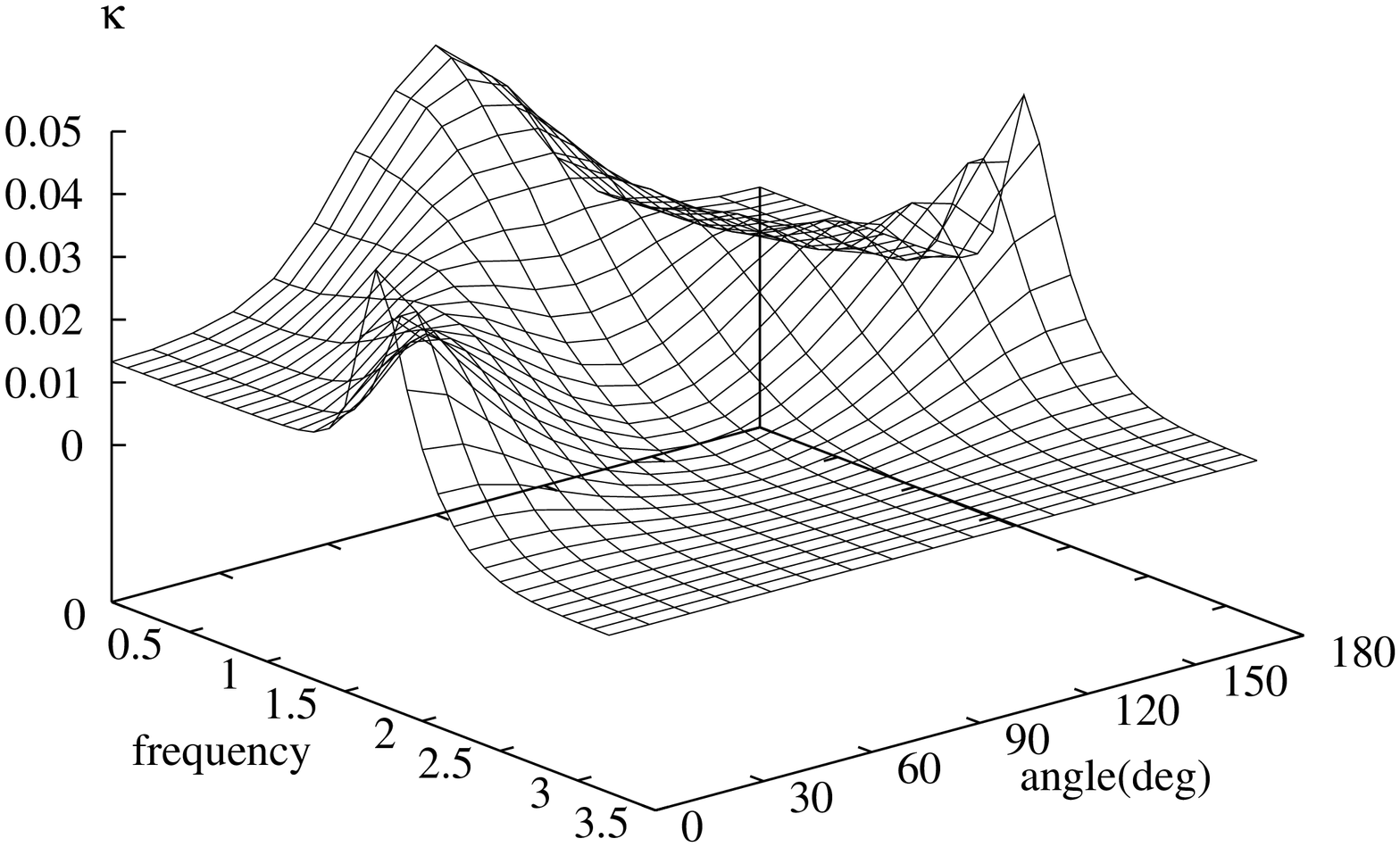}
\caption{An example of the function $\widetilde{\kappa}(R_{e},\theta,\widetilde{\nu})$
in the computations for a spherical maser with seed radiation due to an
isotropic background ($\widetilde{I}_{0}=10^{-9}$). In the computations for this
example, $R/\Gamma=52.$}
\end{figure}

\begin{figure}
\epsscale{1.0}
\includegraphics[angle=270,width=1.0\textwidth]{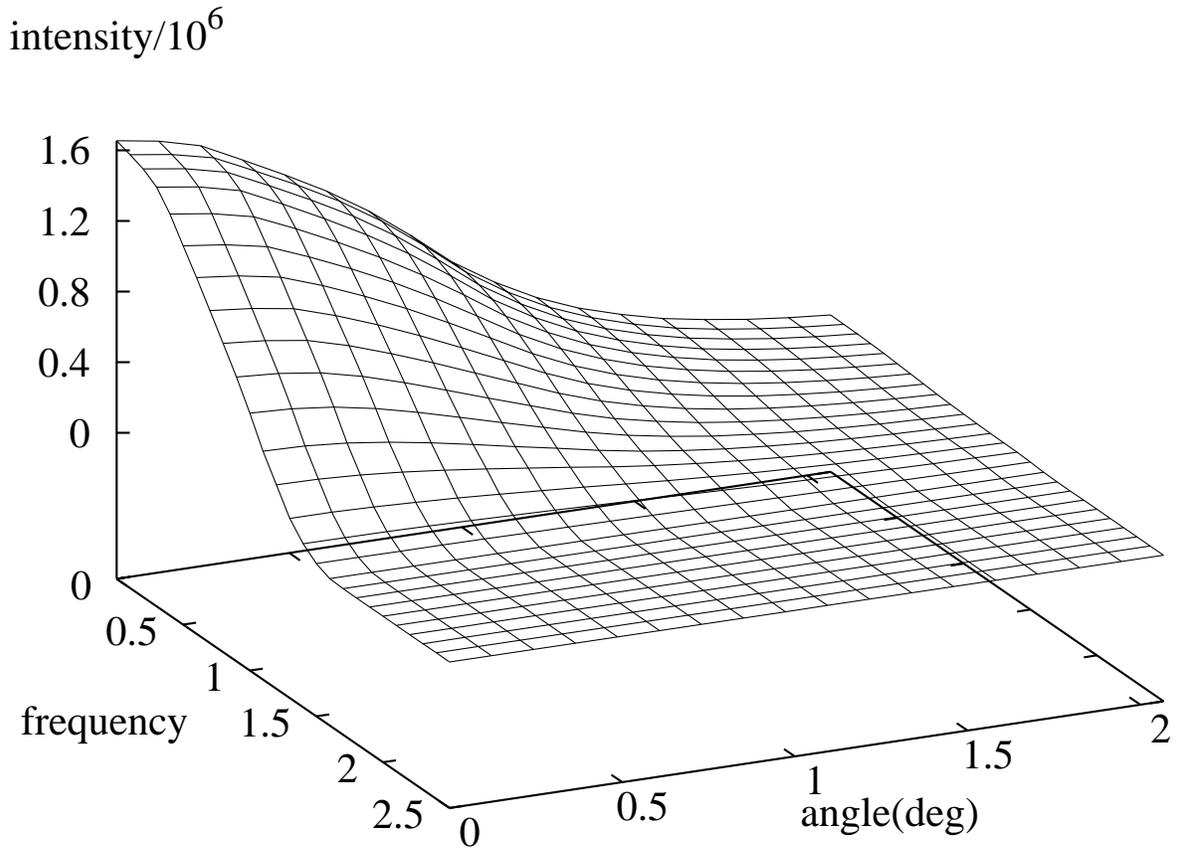}
\caption{An example of the intensity $\widetilde{I}(R_{e}
,\theta,\widetilde{\nu})$ at the surface in
the computations for a spherical maser with seed radiation due to an isotropic
background ($\widetilde{I}_{0}=10^{-9}$). In the computations for this example,
$R/\Gamma=52.$ Note that numerical values on the intensity axis should be multiplied by $10^{6}.$}
\end{figure}

\begin{figure}
\epsscale{1.0}
\plotone{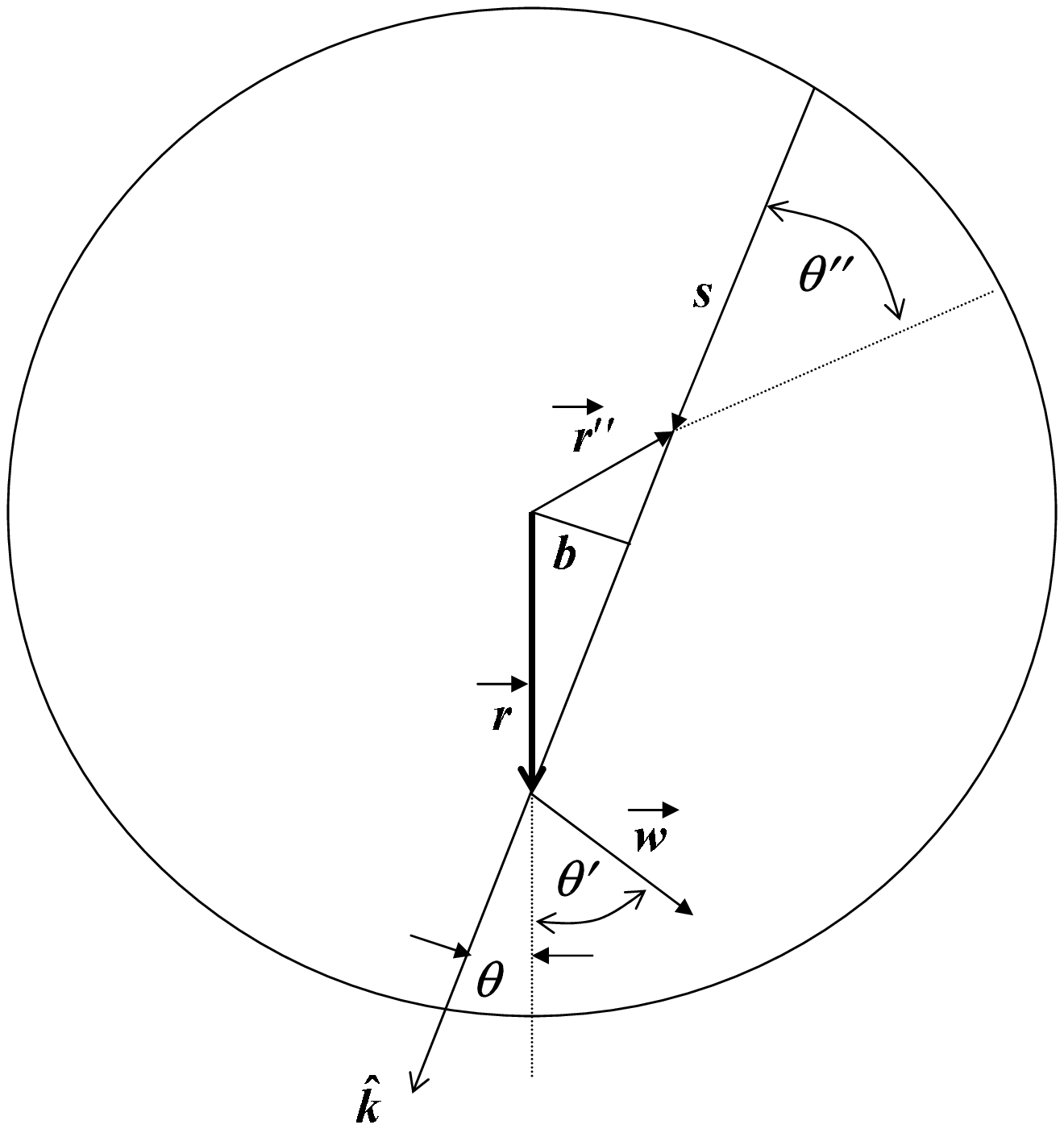}
\caption{Diagram of the geometry.}
\end{figure}


\begin{thebibliography}{}

\bibitem[]{}Anderson, N., \& Watson, W.D. 1993 \apj, 407, 620

\bibitem[]{}Bettweiser, E.V. 1974 A\&A, 50, 231

\bibitem[]{}Bettweiser, E.V.\ \& Kegel, W.H. 1976 A\&A, 37, 291

\bibitem[]{}Emmering, R.T., \& Watson, W.D. 1994 \apj, 424, 991

\bibitem[]{}Goldreich, P., \& Keeley, D.A. 1972 \apj, 174, 517

\bibitem[]{}Goldreich, P., \& Kwan, J.Y. 1974 \apj, 190, 27

\bibitem[]{}Gwinn, C.R. 1994, \apj, 429, 253

\bibitem[]{}Litvak, M. 1971 \apj, 170, 71

\bibitem[]{}Moscadelli, L., Menten, K. M., Walmsley, C. M., \& Reid, M. J. 2003 \apj,
583, 776

\bibitem[]{}Nedoluha, G.E., \& Watson, W.D. 1991 \apj, 367, L63

\bibitem[]{}Neufeld, D.N. 1992, \apj, 393, L37

\bibitem[]{}Wallin, B.K., \&\ Watson, W.D. 1997 \apj, 476, 685

\bibitem[]{}Watson, W.D., Sarma, A.P., \& Singleton, M.S. 2002 \apj, 570, L37
\end{thebibliography}
\end{document}